# A novel modular neutron long counter for (α,n) reactions

**N Mont-Geli**[1,3], **A Tarifeño-Saldivia**[1], **F Calviño**[1], **M Pallàs**[1] **and J L Tain**[2]

[1] Institut de Tècniques Energètiques, Universitat Politècnica de Catalunya, Diagonal 647, 08028 Barcelona, Spain

[2] Instituto de Física Corpuscular, Catedrático José Beltrán 2, E-46980 Paterna, València, Spain

[3] nil.mont@upc.edu

**Abstract**. In this work, we present the design of a new modular and transportable neutron detector for (α,n) reactions. The detector is based on the use of several $^3$He-filled proportional counters embedded in high density polyethylene. In order to provide the detector with a response independent of the neutron energy, a flat response, an innovative design methodology has been applied. The method is based on the optimization of the counter's contribution to the total efficiency by using thermal neutron absorbers. This allows to obtain flat responses up to 10 MeV. The characterization of a proof-of-concept prototype detector is presented. The perspectives for using this new detector for (α,n) reactions measurements are discussed.

## 1. Introduction

*1.1. Physical case*

The production yields and energy spectra from (α,n) reactions are a key magnitude in different research areas. Currently, the accuracy in the calculation of these magnitudes is very constrained by the lack of accurate nuclear data and evaluations.

In nuclear astrophysics, (α,n) reactions play an important role during the α-particle capture process (the α-process) [1]. In fact, the uncertainties on the calculated (α,n) reaction rates have an impact of up to several order of magnitude on the predicted abundance of the light heavy elements (30 < Z < 45) [2]. Another important physical case is the role of the reactions $^{13}$C(α,n)$^{16}$O and $^{22}$Ne(α,n)$^{25}$Mg as a source of neutrons for the slow neutron capture nucleosynthesis process (the s-process) [3].

In underground laboratories, (α,n) reactions are one of the main sources of neutron background, which is an important issue for dark matter and neutrino experiments [4,5]. The uncertainties on the (α,n) reaction cross-sections are a limiting factor when assessing the total background of such kind of experiments [5,6].

*1.2. Neutron long counters for (α,n) reactions*

One of the most common techniques used in (α,n) reactions measurements are the so-called long counters [7], which consist of several gaseous proportional counters embedded in a neutron energy moderator material. A response independent of the neutron energy, a flat response, is required. To quantify this characteristic the so-called flatness factor, which has proved to be useful in previous works [8], is introduced. It is defined as the quotient between the maximum and the minimum efficiencies inside a specific energy range.

In order to improve the accuracy of (α,n) nuclear data, in recent years new long counters have been developed. The main characteristics of the most recent long counters designed for (α,n) reactions are shown.

**Table 1.** Neutron long counters for (α,n) reactions designed in the last ten years.

|  | Nº of counters | Energy range | Average efficiency | Flatness factor |
|---|---|---|---|---|
| Falahat et al [9] | 20 ($^3$He) | (0.3, 0.9) MeV | 55% | 1.6 - 1.7 |
| HabaNERO[a] [10] | 36 ($^3$He) + 44 (BF$_3$) | (0.1, 20) MeV | 22% | 1.23 |
| BB [11] | 18 (BF$_3$) | (0.01, 10) MeV | 3% | 1.32 |

[a] This detector is under development.

## 2. MiniBELEN detector

In this work we report a new modular and transportable neutron long counter (MiniBELEN) based on the same materials and instrumentation than the BEta deLayEd Neutrons (BELEN) long counter [12]. Modularity means that several independent high-density polyethylene (HDPE) blocks are used to form the moderation material matrix. $^3$He-filled proportional counters are used to detect neutrons. They are placed inside the central hole of 7x10x70 cm$^3$ HDPE blocks such as the ones in figure 1. Additional blocks with variable dimensions are used in order to complete the whole detector.

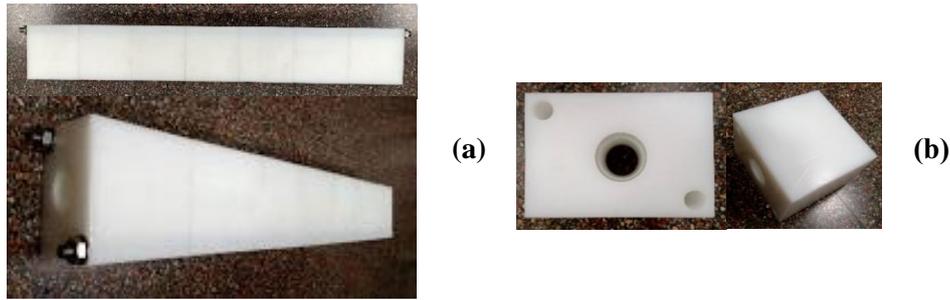

**Figure 1.** HDPE blocks with a central hole for the proportional counters (a). They consist of seven smaller blocks (b) assembled using two stainless steel rods.

### 2.1. Monte Carlo calculations

Monte Carlo calculations of the neutron detection efficiency using a GEANT4 based code are performed in order to find as flat as possible configurations of MiniBELEN with an average efficiency as high as possible. Other important considerations taken into account during the design process are the compactness of the HDPE matrix and the fact that a central hole large enough is required in order to allow the insertion of a 67 mm diameter cylindrical germanium detector.

## 3. Composition method

A traditional design methodology for long counters is the optimization of the radii of the detector rings [8,9,11,13]. In MiniBELEN, due to the modular structure, an alternative approach is required. Therefore, the so-called composition method has been developed and used.

The composition method is based on weighting the contribution of each ring, $\varepsilon_i(E_n)$, to the total detection efficiency, $\varepsilon(E_n)$, using the so-called composition functions, $f_i(E_n)$:

$$\varepsilon(E_n) = \sum_i f_i(E_n) \cdot \varepsilon_i(E_n) \quad (3)$$

The composition functions are physically implemented using cadmium filters (see figure 2) covering partially the active region of the $^3$He tubes in such way that neutrons of any energy that are moderated to energies below 0.4 eV are absorbed. Therefore, the composition functions are nearly independent of the initial neutron energy. The amount of active region covered with cadmium is equal to the value of these functions.

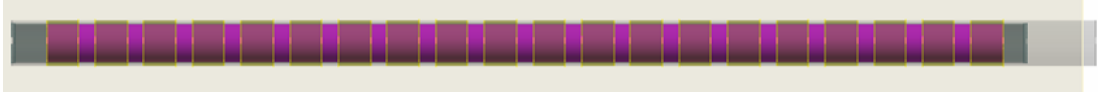

**Figure 2.** GEANT 4 geometry of a $^3$He tube partially covered with a number of cadmium annuli (2 cm length and 0.5 mm thickness) in order to reproduce the composition functions. Cadmium coverage range includes the whole active $^3$He region.

## 4. MiniBELEN configurations

Several configurations have been studied and using the composition method those ones with an optimal flatness and average efficiency in the range 0 up to 10 MeV have been selected. The gas pressure of the $^3$He-filled proportional counters is, generally, 10 atm but a 4 atm and a 20 atm tubes are also used in the outer ring of each configuration.

Geometries are shown in figure 3 and its dimensions are described in table 2. There are two distinguishable regions. The first one is the so-called *detector core* and determines the flatness of the response. The second one is the *reflection region*, which are used in order to increase the efficiency due to the backscattered (reflected) neutrons.

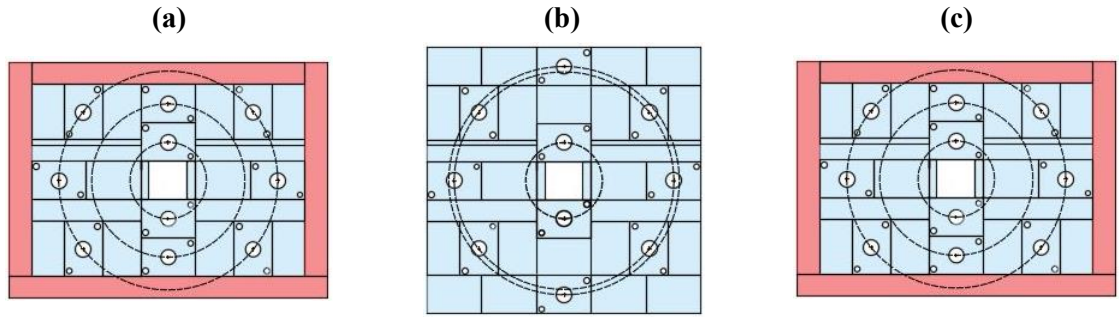

**Figure 3.** MiniBELEN-10A (a), MiniBELEN-10B (b) and MiniBELEN-12 (c). MiniBELEN-10A (a) and MiniBELEN-12 (c) are externally covered with 4 cm width reflectors (red). In MiniBELEN-10B (b). Rings are shown in dotted lines.

**Table 2.** Dimensions of MiniBELEN.

|  | MiniBELEN-10A | MiniBELEN-10B | MiniBELEN-12 |
|---|---|---|---|
| **Core** | 50x35x70 cm$^3$ | 50x49x70 cm$^3$ | 50x35x70 cm$^3$ |
| **Core + Reflectors** | 58x43x70 cm$^3$ | 50x49x70 cm$^3$ | 58x43x70 cm$^3$ |

The amount of cadmium depends on the configuration. While it is used in the first and the second ring of MiniBELEN-10A 10A ($f_1 = 0.4$ and $f_2 = 0.8$) and MiniBELEN-12 ($f_1 = 0.4$ and $f_2 = 0.875$), in MiniBELEN-10B it only used in the inner ring ($f_1 = 0.325$). The neutron detection efficiencies are described in table 3 and figure 4.

**Table 3.** Average efficiency and flatness factor.

| Energy range | MiniBELEN-10A | | MiniBELEN-10B | | MiniBELEN-12 | |
|---|---|---|---|---|---|---|
|  | $\varepsilon_{av}$ | F | $\varepsilon_{av}$ | F | $\varepsilon_{av}$ | F |
| 0 – 5 MeV | 7.097(6) | 1.092(7) | 5.336(6) | 1.189(7) | 8.501(8) | 1.125(5) |
| 0 – 8 MeV | 6.988(5) | 1.143(7) | 5.415(6) | 1.189(7) | 8.435(6) | 1.129(5) |
| 0 – 10 MeV | 6.797(5) | 1.316(7) | 5.336(6) | 1.197(7) | 8.232(6) | 1.285(5) |

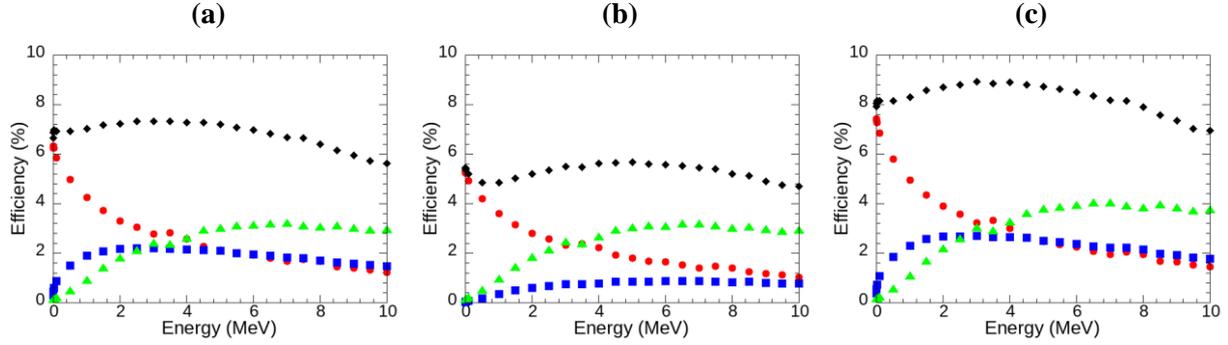

**Figure 4.** Energetic dependence of the response of MiniBELEN-10A (a), MiniBELEN-10B (b) and MiniBELEN-12 (c). Black rhombus: total. Red circles: ring 1. Green triangles: ring 2. Blue squares: ring 3. Error bars are smaller than the size of the data-points. Relative errors are below 0.5 %.

## 5. Experimental characterization of a proof-of-concept prototype

A proof-of-concept prototype of MiniBELEN have been experimentally characterized using $^{252}$Cf spontaneous fission neutron sources and the neutron multiplicity counting (NMC) technique [14] in order to check the performance of the Monte Carlo code. As it has been proved recently [15], NMC allows to determine $^{252}$Cf spontaneous fission neutron yields with metrological precision.

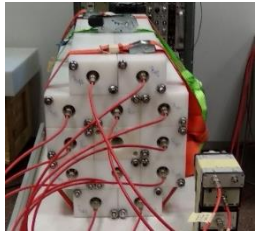

**Figure 5.** Proof-of-concept prototype of the MiniBELEN detector used for the experimental characterization with $^{252}$Cf sources.

In the approximation where the detection efficiency is the same for prompt and delayed neutrons, it has been shown that for a point-like source with negligible (α,n) emission, such as $^{252}$Cf, the neutron detection efficiency can be calculated as [14,16]:

$$\varepsilon_{NMC} = 2f_d \frac{\nu_1 + \nu_d}{\nu_2} \frac{D}{S} \qquad (4)$$

Where $\varepsilon_{NMC}$ is the experimental neutron detection efficiency, $\nu_1$ and $\nu_2$ are, respectively, the first and the second moments of the prompt $^{252}$Cf neutron multiplicity distribution, $\nu_d$ is the average number of emitted delayed neutrons, $S$ is the neutron detection rate (the Singles rate), $D$ is the two-correlated neutrons detection rate (the Doubles rate) and $f_d$ is the so-called gate factor. Using digital electronics, it is trivial to set $f_d = 1$. In order to compare the experimental efficiency and the Monte Carlo calculated efficiency ($\varepsilon_{MC}$), the $R$ parameter is defined as the quotient $\varepsilon_{NMC}/\varepsilon_{MC}$. The $^{252}$Cf spontaneous fission neutron spectrum used in the Monte Carlo calculations is found in reference [17].

### 5.1. Results

Measurements have been carried out at the Technical University of Catalonia (UPC, Barcelona, Spain) and at the Canfranc Underground Laboratory (LSC, Canfranc-Estación, Spain). Three different sources have been used at LSC. The so-called LSC-3 source is the same used at UPC. Background rates were also measured in both locations and subtracted from the $^{252}$Cf rates. Results are shown in figure 6, where two uncertainties are reported: the first ones (short bars) are only statistical and the second one includes the effect of the known HDPE density (0.94 – 0.97 g/cm$^3$). Calculations have been done using 0.95 g/cm$^3$.

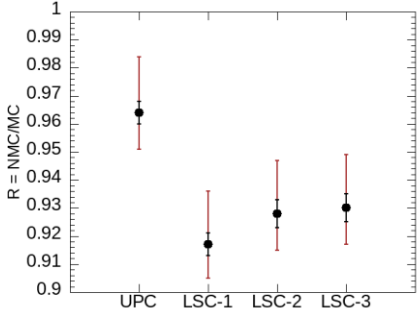

**Figure 6.** Background corrected experimental efficiency normalized to the Monte Carlo calculated efficiency. Shor error bars are statistical. Large error bars include the effect of the poorly known HDPE density.

As it can be seen in figure 6, the efficiency is systematically overestimated by the Monte Carlo code. This is not a surprising result since the same phenomenon is observed in other works with similar setups [9][11][18]. The exact origin of it is unknown. A combination of several factors, such as the experimental uncertainties of the neutron scattering cross-sections or the poorly known HDPE density, could explain it.

Another interesting result is the fact that values of $R$ computed from the LSC measurements are lower than the value compute from UPC data. Because the background rates measured at LSC are negligible, it is reasonable to think that this difference is caused by an inaccurate determination of the UPC background rates due to its variation during the nearly two days measurement.

## 6. MiniBELEN (α,n) reactions

In this section, the performance of the proposed configurations in (α,n) measurements will be discussed using two well-known reactions: $^{27}Al(α,n)^{30}P$ and $^{13}C(α,n)^{16}O$. The (α,n) neutrons detection efficiency is calculated as:

$$\varepsilon_{(\alpha,n)}(E_\alpha) = \int_0^{E_\alpha} F(E)\varepsilon(E)dE \qquad (5)$$

Where $\varepsilon(E)$ is the neutron detection efficiency at energy $E$, $E_\alpha$ is the energy of the incident α particle and $F(E)$ is the (α,n) emission spectrum of this reaction. Monte Carlo calculated neutron spectra [19] are used.

The (α,n) detection efficiencies, normalized to the average efficiency at $E_n$ = 10 MeV, are shown in figure 7. It can be seen how the dependence of $\varepsilon_{(\alpha,n)}$ on the incident α energy is much greater in MiniBELEN-10B than in MiniBELEN-10A or MiniBELEN-12. It is a reasonable and expectable result because up to 8 MeV these configurations are flatter and both (α,n) emission spectra presents peaks below this value.

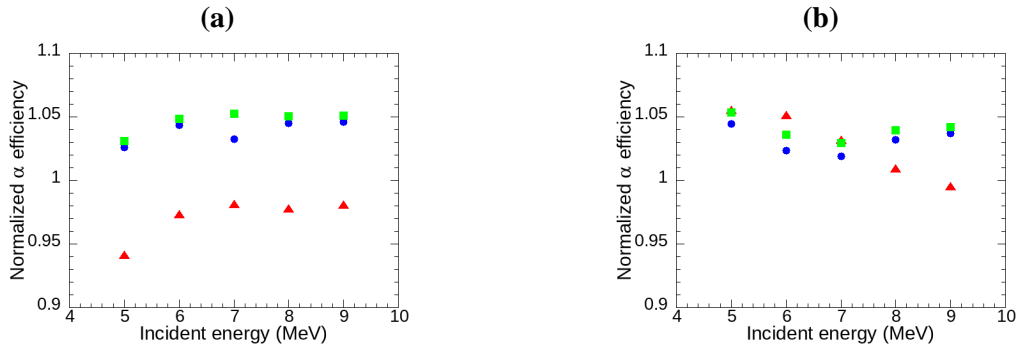

**Figure 7.** Normalized (α,n) detection efficiency for $^{27}Al$ (a) and $^{13}C$ (b) of the three MiniBELEN configurations: 10A (blue circles), 10B (red triangles) and 12 (green squares). Error bars are smaller than the size of the data-points.

## 7. Conclusions

Three configurations of MiniBELEN with detection efficiencies flat up to 10 MeV have been presented. The key issue in the design has been the use of the so-called composition method with cadmium filters. It has been found that the Monte Carlo simulation overestimates the efficiency but this result has been found in works using similar setups. A detailed study of the HDPE density is required in order to check its effect on the overestimation.

The three MiniBELEN configurations reported in this work can be built using, approximately, the same materials. This is one of the great advantages of using a modular detector. The advantage of each one has been discussed in section 6. It has been shown that the neutron emission spectrum of each reaction is a key issue in order to decide which configuration should be used.

**Acknowledgments**

This work has been supported by the Spanish Ministerio de Economía y Competitividad under grants FPA2017-83946-C2-1 & C2-2 and PID2019-104714GB-C21 & C22.